\documentclass{eptcs}

\title{An Improved Algorithm for Generating Database Transactions from
  Relational Algebra Specifications}

\author{Daniel J.\ Dougherty
\institute{Worcester Polytechnic Institute \\
Worcester, MA, USA, 01609}
\email{dd@wpi.edu}
}
\def\titlerunning{An Improved Algorithm for Generating Database Transactions}
\def\authorrunning{Daniel J. Dougherty}

\usepackage{pdfsync}  
\usepackage{pslatex}   

\usepackage{amsmath}
\usepackage{amsfonts}
\usepackage{amssymb}
\usepackage{latexsym}  

\usepackage[mathcal]{euscript}
\usepackage{ifthen}   
\usepackage{ifsym}
\usepackage{xspace}    
\usepackage{epsfig}
\usepackage{verbatim}  
\usepackage{stmaryrd}
\usepackage{times}
\usepackage{xspace}
\usepackage{eepic}
\usepackage{epsfig}
\usepackage{prooftree}
\usepackage{pifont}
\usepackage{breakurl}
\usepackage{verbatim}  

\usepackage{color}
\usepackage{graphicx}
\usepackage{pgf} \usepackage{tikz} \usetikzlibrary{arrows,automata,topaths}

\newcount\timehh\newcount\timemm \timehh=\time 
\divide\timehh by 60 \timemm=\time
\newcommand{\timestamp}{ 
  {\protect\small\sl\today\ -- 
    \ifnum\timehh<10 0\fi\number\timehh\,:\,
    \ifnum\timemm<10 0\fi\number\timemm}}

\usepackage{amsthm}
\theoremstyle{plain} 
\newtheorem{theorem}{Theorem}
\newtheorem{lemma}[theorem]{Lemma}

\newtheorem*{theorem-un}{Theorem}
\newtheorem*{lemma-un}{Lemma}
\newtheorem*{proposition-un}{Proposition}
\newtheorem*{corollary-un}{Corollary}

\theoremstyle{definition} 
\newtheorem{definition}[theorem]{Definition}
\newtheorem*{definition-un}{Definition}

\newtheorem*{notation-un}{Notation}

\newtheorem*{remark-un}{Remark}

\theoremstyle{remark} 

\newtheorem*{example-un}{Example}

\newtheorem*{examples-un}{Examples}

\usepackage{slatex}
\defschememathescape{$} 

\setkeyword{void match then in foreach choose iterate until elseif}
\setkeyword{pred fact sig not no set lone some none one all implies}
\setconstant{+ & ~ -> - . ^ !=}
\setspecialsymbol{FAIL}{{\sc fail}}
\setspecialsymbol{:}{{\textrm :}}
\setspecialsymbol{->}{$\rightarrow$}

\setconstant{var rel none expr binop unop}
\setconstant{formula elemFormula compFormula quantFormula}

\long\def\ignore#1{\relax}

\def\refdef#1{Definition~\ref{#1}}

\def\refthm#1{Theorem~\ref{#1}}

\renewcommand{\phi}{\varphi}

\renewcommand{\to}{\!\rightarrow\!}

\newcommand{\st}{\ensuremath{\; . \;}\xspace}

\newcommand{\alchemyapi}[1]{{\emph{#1}}}

\newcommand{\FAIL}{{\sc fail}}

\newcommand{\tuple}[1]{\mbox{$\langle#1\rangle$}}

\newcommand{\inst}{\ensuremath{I}\xspace}

\newcommand{\state}{\ensuremath{\mathsf{State}}\xspace}

\newcommand{\db}{\ensuremath{I}\xspace}
\newcommand{\node}{\ensuremath{\db}\xspace}

\newcommand{\alconv}{\ensuremath{\sim}\xspace}
\newcommand{\aldiff}{\ensuremath{-}\xspace}

\newcommand{\mthnone}{\ensuremath{\emptyset}\xspace}

\newcommand{\mthunion}{\ensuremath{\cup}\xspace}
\newcommand{\mthint}{\ensuremath{\cap}\xspace}

\newcommand{\mthpair}[2]{\ensuremath{\langle {#1}, {#2} \rangle}}

\newcommand{\mthtc}[1]{{#1}^{*}\xspace}

\newcommand{\sem}[1]{ \llbracket {#1} \rrbracket}
\newcommand{\semenv}[1]{\sem{#1}_{\env}}

\newcommand{\cmd}[1]{\mathsf{code}({#1})}

\newcommand{\env}{\ensuremath{\eta}\xspace}

\newcommand{\thealg}{\ensuremath{\mathbb{A}_{p}}\xspace}
\newcommand{\helperalg}{\ensuremath{\mathbb{B}_{p}}\xspace}
\newcommand{\eval}{\mathbb{E}\mathrm{val}}
\newcommand{\updates}{\ensuremath{\mathsf{Updates}}\xspace}

\newcommand{\tb}{\ensuremath{\qquad}}

\newcommand{\alloyin}{\textbf{ in }}
\newcommand{\alloynotin}{\textbf{ not in }}

\newcommand{\notationNote}{For consistency with the
  presentation and analysis of the algorithms below, we use standard
  mathematical notation in two places where Alloy uses ASCII notation:
  $\cup$ is ``+'' in Alloy, $\cap$ is ``\&''.}

\begin{document} \maketitle
\begin{abstract}
  Alloy is a lightweight modeling formalism based on relational algebra.
  In prior work with Fisler, Giannakopoulos, Krishnamurthi, and Yoo, we
  have presented a tool, Alchemy, that compiles Alloy specifications
  into implementations that execute against persistent databases.  The
  foundation of Alchemy is an algorithm for rewriting relational algebra
  formulas into code for database transactions.  In this paper we report
  on recent progress in improving the robustness and efficiency of this
  transformation.
\end{abstract}

\section{Introduction}
Alloy~\cite{dnj:sw-abstr} is a popular  modeling language that
implements the lightweight formal methods
philosophy~\cite{jw:lightweight-fm}.  Its expressive power is that of
first-order logic extended with transitive closure, and its syntax,
based on relational algebra, is strongly influenced by object modeling
notations.
The language is accompanied by the Alloy Analyzer: the analyzer builds
models (or ``instances'') for a specification using SAT-solving
techniques.  Users can employ a graphical browser to explore instances
and counter-examples to claims.

Having written an Alloy specification, the user must then write the
corresponding code by hand; consequently there are no formal guarantees
that the resulting code has any relationship to the specification.  The
\emph{Alchemy} project addresses this issue.  Alchemy is a tool under
active development
~\cite{kdfy:alchemy-trans-alloy-spec-impl,gdfk:op-sem-alloy} at
Worcester Polytechnic Institute and Brown University, by Kathi Fisler,
Shriram Krishnamurthi, and the author, with our students Theo
Giannakopoulos and Daniel Yoo, that compiles Alloy specifications into
libraries of database operations.  This is not a straightforward
enterprise since, in contrast to Z \cite{jms:z-book} and B
\cite{jra:b-book}, where a notion of state machine is built into the
language, Alloy does not have a native machine model.

Alchemy opens up a new way of working with Alloy specifications: as
declarative notations for imperative programs.  In this way Alloy models
support a novel kind of rule-based programming, in which
underspecification is a central aspect of program design.

In this note we report on recent progress in improving the process of
generating imperative code for declarative specifications in a language
like Alloy.  This paper is a companion to \cite{gdfk:op-sem-alloy},
which developed a better semantic foundation for interpreting Alloy
predicates as operations.  With this better foundation we are able to
generate code for a wider class of predicates than that treated
in~\cite{kdfy:alchemy-trans-alloy-spec-impl} and also prove a more
robust correctness theorem relating the imperative code to the original
specification.

\section{Alloy and Alchemy}


{Some of the material in this expository section is taken
  from~\cite{kdfy:alchemy-trans-alloy-spec-impl}.

\subsection{An overview of Alloy}
An excellent introduction to Alloy is Daniel Jackson's
book~\cite{dnj:sw-abstr}.  Here we start with an informal introduction
to Alloy syntax and semantics via an example.  The example is a homework
submission and grading system, shown in
Figure~\ref{fig:alloy:running-eg}.  In this
system, students may submit work in pairs.  The gradebook stores the grade
for each student on each submission.  Students may be added to or
deleted from the system at any time, as they enroll in or drop the
course.

\begin{figure}
\begin{schemedisplay}
     sig Submission {}
     sig Grade {}
     sig Student {}

     sig Course {
       roster :  set Student, 
       work : roster -> Submission,
       gradebook : work -> lone Grade }

     pred Enroll (c, c' : Course, sNew : Student) {
        c'$$.roster = c.roster $\cup$ sNew and
        c'$$.work[sNew] = $\emptyset$ }

     pred Drop (c, c' : Course, s: Student) {
       s $\alloynotin$ c'$$.roster }

     pred SubmitForPair (c, c' : Course, s1, s2 : Student,
                         bNew : Submission) {
       // $\mbox{pre-condition}$
       s1 $\alloyin$ c.roster and s2 $\alloyin$ c.roster and  
       // $\mbox{update}$
       c'$$.work = c.work $\cup$ <s1, bNew> $\cup$ <s2,  bNew> and
       // $\mbox{frame condition}$
       c'$$.gradebook = c.gradebook }

     pred AssignGrade (c, c' : Course, s : Student, 
                       b : Submission, g : Grade) {
       c'$$.gradebook $\alloyin$ c.gradebook $\cup$ <s,  b, g> and
       c'$$.roster = c.roster }

     fact SameGradeForPair {
       all c : Course, s1, s2 : Student, b : Submission | 
          b in (c.work[s1] & c.work[s2]) implies
             c.gradebook[s1][b] = c.gradebook[s2][b] }
\end{schemedisplay}
\caption{Alloy specification of a gradebook.}\hrule
\label{fig:alloy:running-eg}
\end{figure}

The system's data model centers around a course, which has three
fields: a roster (set of students), submitted work (relation from
enrolled students to submissions), and a gradebook.  Alloy uses %
\textbf{sig}natures %
to capture the sets and relations that comprise a
data model.  Each \scheme{sig} (\scheme{Submission}, etc.)  defines a
unary relation.  The elements of these relations are called
\emph{atoms}; the type of each atom is its containing relation.

Fields of signatures define additional relations.  The \scheme{sig}
for
\scheme{Course}, for example, declares \scheme{roster} to be a
relation on \scheme{Course}$\times$\scheme{Student}.  Similarly, the
relation \scheme{work} is of type
\scheme{Course}$\times$\scheme{Student}$\times$\scheme{Submission},
but with the projection on \scheme{Course} and \scheme{Student}
restricted to pairs in the \scheme{roster} relation.  The
\scheme{lone} annotation on \scheme{gradebook} allows at most
one grade per submission.

The \textbf{pred}icates (\scheme{Enroll}, etc.)\ capture the actions
supported in the system.  The predicates follow a standard Alloy idiom
for stateful operations: each has parameters for the pre- and
post-states of the operation (\scheme{c} and \scheme{c'}, respectively),
with the intended interpretation that latter reflects a change applied
to the former.  Alloy \textbf{fact}s (such as \scheme{SameGradeForPair})
capture invariants on the models.  This particular fact states that
students who submit joint work get the same grade.

An important aspect of Alloy is that \emph{everything is a relation.}
In particular sets are viewed as unary relations, and individual atoms
are viewed as singleton unary relations.  As a consequence the
$\alloyin$ operator does double-duty: it is interpreted formally as
subset, but also stands in for the ``element-of'' relation, in the sense
that if---intuitively---$a$ is an atom that is an element of a set $r$,
this is expressed in Alloy as $a \alloyin r$, since $a$ is formally a
(singleton) set.

The Alloy semantics defines a set of models for the signatures and
facts.  Operators over sets and relations have their usual semantics:
$\cup$ (union), $\cap$ (intersection), $\langle \;,\; \rangle$
(tupling), and \scheme{.}  (join).\footnote{\notationNote}  As noted above, $\alloyin$ denotes
subset and is also used to encode membership.  Square brackets provide a
convenient syntactic sugar for certain joins: $e2[e1]$ is equivalent to
$e1.e2$.  The following relations constitute a model under the Alloy
semantics.
\begin{schemedisplay}
     Student = {Harry, Meg}
     Submission = {hwk1}
     Grade = {A, $A-$, B+, B}
     Course = {c0, c1}
     roster = ($\tuple{\scheme|c0, Harry|}$, $\tuple{\scheme|c1, Harry|}$, $\tuple{\scheme|c1, Meg|}$)
     work = {$\tuple{\scheme|c1, Harry, hwk1|}$}
     gradebook = {$\tuple{\scheme|c1, Harry, hwk1, A-|}$}
\end{schemedisplay}

\noindent
A model of a predicate also associates each
predicate parameter with an atom in the model such that the predicate
body holds.  The above set of relations models the \scheme{Enroll}
predicate under bindings \scheme{c} =
\scheme{c0}, \scheme{c'} = \scheme{c1} and \scheme{sNew} =
\scheme{Meg}.  A model may include tuples beyond those required to
satisfy a predicate: the \scheme{Enroll} predicate does not
constrain the \scheme{work} relation for pre-existing students, so
the appearance of tuple \tuple{\scheme|c1, Harry, hwk1|} in the
\scheme{work} relation is semantically acceptable.

The reader may want to check that the relations shown do not happen to
model the predicate \scheme{SubmitForPair}, in the sense that no
bindings for $c, c', s1, s2, and bNew$ make the body of
\scheme{SubmitForPair} true.    Under
\scheme{c} = \scheme{c0} and \scheme{c'} = \scheme{c1}, for example, the requirement
\scheme{c'}\scheme{.gradebook = c.gradebook} fails because the gradebook
starting from \scheme{c'} has one tuple while that starting from
\scheme{c} has none.  The requirement on \scheme{work} also fails.
Similar inconsistencies contradict other possible bindings for
\scheme{c} and \scheme{c'}.
   

\subsection{An overview of Alchemy}

We illustrate Alchemy in the context of  the gradebook specification from
Figure~\ref{fig:alloy:running-eg}.    Alchemy creates a database table for
each relation (e.g., \scheme{Submission}, \scheme{roster}), a procedure
for each predicate (e.g., \scheme{Enroll}), and a function for creating
new elements of each atomic signature (e.g., \scheme{CreateSubmission}).
 A sample session using
Alchemy might proceed as follows.  We create a course with two students using
the following command sequence:
\begin{schemedisplay}
     cs311 = CreateCourse("cs311");
     pete = CreateStudent("Pete");
     caitlin = CreateStudent("Caitlin");
     Enroll(cs311, pete);
     Enroll(cs311, caitlin)
\end{schemedisplay}
Note that the \scheme{Enroll} function takes only one course-argument,
in contrast to the two in the original Alloy predicate, since the implementation
maintains only a single set of tables over time (the second course
parameter in the predicate corresponds to the resulting updated table).
Executing the
\alchemyapi{Enroll} function adds the pairs 
\tuple{\scheme|"cs311"|,~\scheme|"Pete"|} and
\tuple{\scheme|"cs311"|,~\scheme|"Caitlin"|} to the
\scheme{roster} table.  The second clause of the \scheme{Enroll}
specification guarantees that the \scheme{work} table will not have
entries for either student. 


Next, we submit a new homework for
\scheme|"Pete"| and
\scheme|"Caitlin"|:
\begin{schemedisplay}
     hwk1 = CreateSubmission("hwk1");
     SubmitForPair(cs311, pete, caitlin, hwk1)
\end{schemedisplay}
The implementation of \scheme{SubmitForPair} is straightforward
relative to the specification.  It treats the first clause in the
specification as a pre-condition by terminating the computation with
an error if the clause is false in the database at the start of the
function execution.  Next, it adds the \scheme{work} tuples
required in the second (update) clause.  It ensures that the 
\scheme{gradebook} table is unchanged, as required by the third
clause.  

Assigning a grade illustrates the way that Alloy facts constrain
Alchemy's updates:
\begin{schemedisplay}
     gradeA = CreateGrade("A");
     AssignGrade(cs311, pete, hwk1, gradeA)
\end{schemedisplay}
\scheme|AssignGrade| inserts a tuple into the \scheme{gradebook}
relation according to the first clause, and checks that the roster is
unchanged according to the second.  If execution were to stop here,
however, the resulting tables would contradict the
\scheme{SameGradeForPair} invariant (which requires \scheme|"Caitlin"|
to receive the same grade on the joint assignment).  Alchemy determines
that adding the tuple \tuple{\scheme|cs311|, \scheme|Caitlin|,
  \scheme|hwk1|,~\scheme|A|} to \scheme{gradebook} will satisfy both
the predicate body and the \scheme{SameGradeForPair} fact, and executes
this command automatically.  If there is no way to update the database
to respect both the predicate and the fact, Alchemy will raise an
exception.  This could happen, for example, if the first clause in
\scheme{AssignGrade} used $=$instead of $\alloyin$:  in this
case, adding the repairing tuple would violate the predicate body).


\paragraph{Maintaining invariants}
\label{invariants}
Alloy's use of \emph{facts} to constrain possibly-underspecified
predicates offers a powerful lightweight modeling tool.  The {facts} in
an Alloy specification are axioms in the sense that they hold in any
instance for the specification.  We may view the facts as integrity
constraints: they capture the fundamental invariants to be maintained
across all transactions. 
Alchemy will guarantee preservation of all facts as database invariants.
This is akin to the notion of \emph{repair} of database transactions.


\subsection{Formalities}

\paragraph{Alloy specifications}
Formally, the \emph{Alloy specifications} we treat in this paper are
tuples of \emph{signatures}, \emph{predicates,} and \emph{facts.}  In
practice Alloy specifications may also include \emph{assertions} to be
checked by the analyzer, but they do not play a direct role in Alchemy's
code generation so we omit them here.
\begin{itemize}
\item  A signature specifies its type
  name and a set of fields.  Each field has a name and a type
  specification $A_0 \to A_1 \to \dots \to A_n$, where each $A_i$ is
  the type name of some signature.

\item A predicate has a header and
  a body.  The header declares a set of variable names, each with an
  associated signature type name; the body is a formula
  in which the only free variables are defined in the header. 

\item A fact is a closed formula, having the force of an axiom: models
  of a specification are required to satisfy these facts.
Alloy
  permits the user to specify certain constraints on the signatures and
  fields when they are declared, such as ``relation $r$ may have at most
  one tuple.''  These can be alternatively expressed as facts and, for
  simplicity of presentation, we assume this is always done.
\end{itemize}

The following language for expressions and formulas is essentially
equivalent to the Kernel language of Alloy~\cite{dnj:sw-abstr} (modulo
the lexical differences between standard mathematical notation used here and
Alloy's ASCII). 
\begin{schemedisplay}
      expr ::= rel | var | none | expr binop expr | unop expr
      binop ::= $\cup$ | $\cap$ | - | . | $\mthpair{}{}$
      unop ::=  $\sim$| $*$

      formula ::= elemFormula | compFormula | quantFormula
      elemFormula ::= expr $\alloyin$ expr | expr = expr  
      compFormula ::= not formula | formula $\land$ formula | formula $\lor$ formula
      quantFormula ::= $\forall$ var: expr { formula } | $\exists$ var: expr { formula }
\end{schemedisplay}

\paragraph{State-based  specifications}

The elements of an Alloy specification suggest natural implementation
counterparts.  The signatures lay out relations that translate directly
into persistent database schemas.  The facts---those properties that are
meant to hold of all models constructed by Alloy---function as database
integrity constraints.  Finally, under a commonly idiom, certain
predicates in an Alloy specification connote state changes.  It is these
state-based specifications that Alchemy (currently) treats.

The state-transition idiom is a commonly understood convention rather
than a formal notion in Alloy.  To precisely define the class of
specifications that Alchemy treats, we first require some terminology.
Fix a distinguished signature, which we will call \state. %
An \emph{immutable type} is one with no occurrences of the \state
signature.

\begin{quote}
{\em The assumptions Alchemy makes about the specifications it treats
  are:}
  \begin{itemize}
  \item specifications are state-based, and

  \item facts have at most one variable of type \state,  and this
    variable is unprimed and universally quantified.
  \end{itemize}
\end{quote}


  
\paragraph{An operational semantics}
The static semantics of Alloy is based on the class of relational
algebras.  To give an operational semantics for state-based Alloy
specifications, one that takes seriously the reading of predicates as
state-transformers, we pass to the class of transition systems whose
nodes are relational algebras.  We also assume that each state has a
single atom of type \state.   When individual relation algebras are read
as database instances, transitions between states can be viewed as
database update sequences transforming one state to another.
We adopt a constant-domain assumption concerning our transition systems.
Space consideration prohibit us from presenting the motivation and
justification for this (including the explanation why it is not as great
a restriction as it may appear); details are in~\cite{gdfk:op-sem-alloy}.

Since predicates have parameters, the meaning of a predicate is relative
to bindings from variables to values.  It is technically convenient to
assume that for a given specification we identify, for each type, a
universe of possible values of this type.    Then
an \emph{environment} \env is a mapping from typed variables to values.

\begin{definition}[\textbf{Operational semantics of predicates}]
  \label{def:op-sem-pred}

  Let $p$ be a predicate  with the property that %
  $p$ has among its parameters exactly two variables \texttt{s} and
  \texttt{s'} of type \texttt{State}, and
  let \env be an environment.  The meaning $\sem{p}\env$ of
  $p$ under \env is the set of pairs $\langle \node, \node'
  \rangle$ of instances such that
  \begin{itemize}
  \item $\env$ maps the parameters of $p$ into the set of atoms of
    $\node$ (which equals the set of atoms of $\node'$), mapping %
    the unprimed \state parameter to the \state-atom of $\node$ and the
    primed \state parameter to the \state-atom of $\node'$;

  \item   $(\node, \node')$ makes
    the body of $p$ true under the environment $\env$: occurrences of
the \state variable $s$ are interpreted in $\node$, while occurrences of
the \state variable $s'$ are interpreted in $\node'$.
  \end{itemize}
\end{definition}

The meaning of a predicate $p$ is a \emph{set} of transitions because
$p$ can be applied to different nodes, with different bindings of the
parameters of course, but also---and more interestingly---because
predicates typically under-specify actions: different implementations of
a predicate can yield different outcomes $\node'$ on the same input
$\node$.  Any of these should be considered acceptable as long as the
relation between pre- and post-states is described by the predicate.

\section{ Main Result} 
We observed that a predicate $p$ determines a family of binary relations over instances,
parametrized by environments.
 That is, for a given environment \env:
 \begin{align}
   \label{pred-sem}
   \sem{{p}}\env &: {Inst \to  2^{Inst}}.   
 \end{align}
 Now suppose $t$ is a procedure defining a database transaction (so $t$
 is the sort of procedure that a predicate $p$ specifies).  Given an
 instance $\inst$ and an environment \env, $t$ may return a new instance
 $\inst'$, terminate with failure, or may diverge.  None of the
 procedures we describe in this paper will diverge, so we are
 considering procedures $t$ that (under an environment) determine a
 function over instances:
 \begin{align}
   \label{proc-sem}
   \sem{t} \env : Inst \to (Inst + {\FAIL}).
 \end{align}
 Alchemy's job is precisely the following: given predicate $p$, construct a
 procedure $t = \cmd{p}$ such that the semantics of $\cmd{p}$ as given
 in \ref{proc-sem} refines the semantics of $p$ as given in
 \ref{pred-sem},  in the following sense. %

 \begin{theorem}[\textbf{Main theorem}] \label{thm-main} Let $p$ be a
   predicate and let $\cmd{p}$ be any backtracking implementation of the
   algorithm \thealg, given in \refdef{def:thealg} below. %
   Then for each instance \inst and each environment \env

  \begin{enumerate}
  \item %
    $\sem{\cmd{p}}\env$ terminates on \inst;
  \item %
    If there exists any instance $\inst'$ such that $(\inst, \inst')$
    satisfies $p$ under $\env$ then %
    the result of $\sem{\cmd{p}} \env$ is such an $\inst'$. In
    particular in this situation $\sem{\cmd{p}}$ does not return ``failure'' under
    $\env$ on \inst.

  \end{enumerate}
\end{theorem}
\begin{proof}
The proof is given in Section~\ref{correctness}.
\end{proof}
It is worth noting that the task of generating updates from
specification submits to an uninteresting trivial solution, particularly
if we are willing to tolerate partial functions.  Given predicate $p$ we
could define $\cmd{p}$ by:
 \emph{on input $I$, exhaustively generate all possible $I'$; for each one
  test whether $(I, I')$ in $\sem{p}$. If and when such an $I'$ is
  found,  replace $I$ by $I'$.}
Obviously this is a silly algorithm, even though it is ``correct'' in a
formal sense.  Our goal with Alchemy is to write code that is
intuitively reasonable, and still is correct  in the
sense of \refthm{thm-main}.

\section{Code generation}

Suppose we are given an Alloy predicate \textbf{p}.  Alchemy generates code
for a procedure with parameters corresponding to those of
\textbf{p} (without the primed parameter).

As observed above, a crucial aspect of \emph{Alloy} is that it
encourages ``lightweight'' specifications of procedures: the designer is
free to ignore details about the computation that she may consider
inessential.  As a consequence, \emph{Alchemy} must be extremely
flexible: different input instances may require quite different
computations in order to satisfy a specification, yet Alchemy must
generate code that works uniformly across all instances.

The top-level view of how Alchemy generates code for a procedure is as
follows.
\subsection{Outline}
\begin{itemize}
\item 
In \refdef{def:thealg} below we present a construction that, based on
predicate $p$, builds a \emph{non-deterministic procedure} \thealg.

\item The code generated by Alchemy, $\cmd{p}$, is a backtracking
  implementation of \thealg.  Computation paths that do not succeed are
  recognized as such and abandoned, and \thealg is finite-branching, so 
$\cmd{p}$ will always terminate.

\item   If there exists any instance $\inst'$ such that $(\inst, \inst')$
    satisfies $p$ under $\env$ then some branch of \thealg is guaranteed
    to compute such some such instance.
\end{itemize}
\paragraph{Coping with inconsistent predicates}
It is possible for the code for a predicate $p$ to \emph{fail} on a
given database instance $I$, either because the predicate is internally
inconsistent or because no update of $I$ can implement $p$ without
violating the facts.  Alchemy is guaranteed to detect such situations;
we treat predicates as \emph{transactions} that rollback if they cannot
be executed without violating their bodies or a fact.

\subsection{A normal form for predicates}
The general form of an Alloy predicate that  specifies an operation and
that Alchemy treats is
\[
\textsf{pred} \; p(s, s' : \state, \; a_1 : A_1, \dots , a_n: A_n) 
\{  \vec{Q x} \st \beta(\vec{a}, \vec{x}) \}
\]
where  $\vec{Q}$ is a sequence of quantified atoms and $\beta$ is a
quantifier free formula of relational algebra.
Before giving an imperative interpretation of a predicate it is
convenient to massage it into a convenient form.

\paragraph{Skolemization} 
By the classical technique of Skolemization any formula %
$ \vec{Q x} \st \beta(\vec{a}, \vec{x}) $ can be converted into a
universal formula which is satisfiable if and only if $ \vec{Q x} \st
\beta(\vec{a}, \vec{x}) $ is satisfiable.  We exploit this trick in
Alchemy as follows.  Given a predicate $p$ we convert it to a predicate
$p^{\forall}$ whose body is in universal form; this involves
expanding the specification language to include the appropriate Skolem
functions.  Suppose we generate code for $p^{\forall}$ (over the
expanded language).  Then given an original instance \inst we may view it as
an instance $\inst_{+}$ over the enlarged schema, and apply the
generated code  to
obtain an instance $\inst'_{+}$.   We ultimately  return the instance
$\inst'$ that is the reduct of $\inst'_{+}$ to the original schema.
So in what follows  we restrict attention to predicates whose body is a
universal formula.

\vspace*{-5pt}
\paragraph{Incorporating the facts}
Intuitively the facts in a specification comprise a separate set of
constraints on how a predicate may build new instances from old ones.
But by the following simple trick we can avoid treating the facts
separately.  When compiling a predicate to code we take each fact, prime
every occurrence of the \state sig, and add the fact to the body of the
predicate.  The use of primed \state names means that the fact acts as a
post-condition on the predicate.  (Strictly speaking this is only true
under an assumption of ``state-boundedness'' on the form of the facts,
defined in \cite{gdfk:op-sem-alloy}.  The specifics of this syntactic
assumption are irrelevant to the current paper so we omit details.) This
in turn guarantees that any post-instance defined by the predicate will
satisfy the facts.

\newcommand{\thefn}{strictly speaking this is only
  true under an assumption of ``state-boundedness'' on the form of the
  facts, defined in \cite{gdfk:op-sem-alloy}.  The specifics of this
  syntactic assumption are irrelevant to the current paper so we omit
  details.}

The following is a convenient form for formulas.
\begin{definition}[\textbf{Special formulas}]
A \emph{special} formula is a formula in either of the two forms
\[
(e_1 \mthint \dots \mthint e_k) = \mthnone  \qquad \text{or} \qquad 
(e_1 \mthint \dots \mthint e_k) \neq \mthnone  
\]
for $k \geq 1$, with each $e_i$ not containing \mthunion or \mthnone and with converse
applied only to variables and relation names.
\end{definition}

\begin{lemma} \label{thm:special}
  Any quantifier-free formula can be transformed into an equivalent
  Boolean combination of special formulas.
\end{lemma}
\begin{proof}
 It is easy to see that every expression is equivalent to one in
  which the converse operator \alconv applies only to relation names or
  variables.  
  It is easy to see that every expression other than \mthnone\ itself is
  equivalent to one in which the constant \mthnone never appears.
 Because union distributes over the other connectives every
  expression is equivalent to one of the form $e_1 \mthunion \dots
  \mthunion e_n$
  ($n \geq 1$) with each $e_i$ being $\mthunion$-free. 

 We may take any equation $e = f$ and replace it with $(e \alloyin f)
  \land (f \alloyin e)$.  We do this as long as neither $e$ nor $f$ is the
  term $\mthnone$.

Now each basic formula is in one of the forms
\[
  (d_1 \mthunion \dots \mthunion d_m) \alloyin
  (f_1 \mthunion \dots \mthunion f_n)
\qquad \mbox{ or } \qquad
  (d_1 \mthunion \dots \mthunion d_m) \alloynotin
  (f_1 \mthunion \dots \mthunion f_n)
\]
with $n, m \geq 0$, 
where the $d_i$ and the $f_i$ are $\mthunion$-free.
  We may transform the basic formulas
  above into the corresponding forms
  \begin{equation} \label{basic-eq}
    (d_1 \mthunion \dots \mthunion d_m) \aldiff
    (f_1 \mthunion \dots \mthunion f_n)
    = \mthnone, 
\qquad \mbox{ respectively, } \qquad
    (d_1 \mthunion \dots \mthunion d_m) \aldiff
    (f_1 \mthunion \dots \mthunion f_n)
    \neq \mthnone
  \end{equation}
  The first
  equation in \ref{basic-eq} is equivalent, via distributivity of \mthunion over
  \mthint, to the \emph{conjunction} of the equations
\[
d_i  \aldiff  (f_1 \mthunion \dots \mthunion f_n) = \mthnone
\qquad 1 \leq i \leq m
\]
In turn, \emph{each} of these is equivalent to the special formula
\[
(d_i  \aldiff  f_1) \,  \mthint  \dots \mthint (d_i \aldiff f_n) = \mthnone
\]
Similar reasoning shows that each dis-equation as in \ref{basic-eq}
is equivalent to a disjunction of special formulas
 \[
 ( \dots ((d_i  \aldiff  f_1) \aldiff f_2)  \aldiff \dots \aldiff f_n) \neq \mthnone
 \]


\end{proof}



\subsection{Algorithms}
\label{sec:alg}

\paragraph{Bridging the declarative/imperative gap} 
The main procedure \thealg below is generated by an induction that walks the
structure of the formula that is the body of ${p}$. %
There is a natural correspondence between the logical operators in the
predicate and control-flow operators in the generated procedure.  The
disjunctive (logical $\vee$ and $\exists$) constructors in predicates
naturally suggest imperative nondeterminism; this of course results in
\emph{backtracking} in generated code.  Likewise, conjunctive (logical
$\wedge$ and $\forall$) constructors lead naturally to
\emph{sequencing}.  This is natural enough, but a difficulty arises due
to the fact that the logical operators are commutative but
command-sequencing certainly is not.  Indeed, implementing one part of a
predicate can undo the effect achieved by an earlier part.  The solution
is to iterate computation until a fixed-point is reached on the post-state.
So we must be careful to ensure that such an iteration will always halt.

\vspace*{-5pt}
\paragraph{Compiling special formulas to code} \label{special}
Consider for example
the body of the \emph{Drop} predicate in
Figure~\ref{fig:alloy:running-eg}.  There are certainly many ways to
update the data to make this true; for example we could delete all the
tuples in the roster table!  This is not what the specifier had in mind.
But even this silly example points out the need for a principled
approach to update.  We start with the following goal: we attempt to
make a \emph{minimal} set of updates (measured by the number of tuples
inserted or deleted into tables) to the system to satisfy the predicate.

The virtue of special formulas is that they facilitate identifying
minimal updates to make a formula true.  For example the formula $a
\alloyin s'.r$, which, when $a$ is an atom, is to say that $a$ is in the
relation $s'.r$ is equivalent to the formula $a - (s'.r) = \mthnone$.
So suppose $a - (s'.r)  = \mthnone$ is part of
the body of a predicate.    
We evaluate the expression $a - (s'.r) $ in the pre-state and the current
 post-state: if the value of this expression is indeed empty then
there is nothing to do.  If it is not empty then $a$ is not in $s'.r$,
and it is clear what action to take: add $a$ to $s'.r$.  

More generally, when confronted with a special formula
 $e = \mthnone$ we may view any tuples in the current value of $e$ as 
\emph{obstacles to the truth of the formula}.   Then the action
suggested by the formula is clear:  make whatever insertions or
deletions we can to ensure the formula becomes true.  (The presence of
the difference operator means that making an expression empty may
involve insertions.)   The important thing to note is that,
obviously, we may focus exclusively on tuples that are already in the value
of $e$ in attempting to make $e = \mthnone$ in the updated state.  This
is our strategy for doing minimal updates for a predicate.

\vspace*{-5pt}
\paragraph{Inserting and deleting tuples} \label{tuples} %
We have seen that compiling a special formula amounts to orchestrating
the insertion or deletion of individual tuples from the relations
denoted by expressions.  These expressions correspond to database
\emph{views}, and indeed the task of inserting or deleting a tuple from
a view  is an instance of the well-known \emph{view
  update} problem~\cite{BlakeleyLT86,BraganholoDH04}.    Our code
proceeds by a structural induction over the expression: see the
procedures insertTuple and deleteTuple below.

\vspace*{-5pt}
\paragraph{Putting it all together}\label{all-together}
After the preceding discussion the pseudocode for the Alchemy's
translation algorithm should be largely self-explanatory.  For
simplicity in notation we adopt the following conventions.  There are
global variables pre-state and post-state ranging over instances, and a
global variable \updates which keeps a record of the insertions and
deletions done as the algorithm progresses.

We make use of the following function 
$  \eval(e : \mbox{expression}, \;  J, J' : \text{database instances})
$ 
that   returns the set of tuples denoted by expression $e$ under the convention
  that immutable relation-name occurrences are interpreted in $J$
  and mutable relation-name occurrences are interpreted in $J'$.
  The pseudocode given here for procedures \thealg, \helperalg,
  insertTuple, and deleteTuple is directly based on the discussion in
  the previous paragraphs.

\begin{definition}[Algorithm \thealg] \label{def:thealg}
  Let $p$ be a Alloy predicate of the form 
  \begin{align*}
    \textsf{pred} \; \; p(s, s' : \state, \;  a_1 : A_1,  \dots , a_n:
    A_n) \; . \; \{ \forall \vec{x} \st \bigwedge_{i} \bigvee_{j} \sigma_{i,
      j} \}
  \end{align*}
  where each $\sigma_{i,j}$ is a special formula.  The procedure \thealg
  determined by $p$ is as follows.
  Each of \thealg and \helperalg reads the instance \inst globally and
  reads and writes \inst' and \updates globally.

\end{definition}

\begin{quote}
\noindent \textbf{procedure} \thealg (\inst: database instance) \{ \\
\tb initialize poststate \inst'  to be \inst;\\
\tb initialize \updates to be empty; \\
\tb repeat  $\helperalg(a_1: A_1, \dots, a_n : A_n)$ \\
\tb until no change in \updates \\
\} \\
\noindent  \textbf{procedure} $\helperalg(a_1 : A_1, \dots, a_n: A_n)$ \{ \\
\tb for each binding $\vec{b}$ of values in \inst for the variables in
$\vec{a}$: \\
  \tb let $\bigwedge_{i} \bigvee_{j} \bar{\sigma}_{i, j}$
  be the body of $p$ instantiated by $\vec{b}$: \\
  \tb \tb for each conjunct %
  $\bigvee_{j} \bar{\sigma}_{i, j}$ \\
  \tb \tb \tb \tb \textbf{choose} some $\bar{\sigma}_{i,j}$ and 
  realize $\bar{\sigma}_{i,j}$ as follows: \\  
  \tb \tb \tb \tb \text{Case 1:} 
    $\bar{\sigma}_{i,j}$ is of the form 
    $(e_1 \mthint \dots \mthint e_k) = \mthnone$  \\
  \tb \tb \tb \tb set $e \equiv {(e_1 \mthint \dots \mthint e_k)}$    \\
  \tb \tb \tb \tb    for each tuple $t$ in $\eval(e, I, I')$:    \\
  \tb \tb \tb \tb     \tb call \emph{deleteTuple$(t, e, I, I')$}; \\ 
  \tb \tb \tb \tb   \text{Case 2:} 
    $\bar{\sigma}_{i,j}$ is of the form 
    $(e_1 \mthint \dots \mthint e_k) \neq \mthnone$ \\
  \tb \tb \tb \tb    set $e \equiv {(e_1 \mthint \dots \mthint e_k)}$  \\
  \tb \tb \tb \tb   \textbf{choose} some $t$ of the same type as $e$ \\
  \tb \tb \tb \tb    \tb    call \emph{insertTuple(t, e. I, I')} \\
  \tb \tb \tb update \updates accordingly; \\
  \}
\end{quote}

\begin{quote}
\noindent \textbf{procedure} insertTuple($t:$ tuple, \; $e$: expression) \{ \\
\tb  match $e$:  \\
%
  \tb atom $a$: if $a \neq t $ then FAIL else RETURN \\
  \tb immutable relation $r$: if $t \notin r$ then FAIL else RETURN \\
  \tb mutable relation $r$: if $t$ has been previously deleted from  $r$
  then FAIL \\
  \tb \tb \tb else add $t$ to the table $r$ in $J'$ \\
  \tb   $e_1 \mthunion e_2$: \textbf{choose} some $ e_i $ ;  insertTuple($t, e_i$)\\
  \tb $e_1 \mthint e_2$: insertTuple($t, e_1$) ; insertTuple($t, e_2$) \\
  \tb $\sim e$: insertTuple($~t, e)$ \\
 \tb $\mthpair{e_1}{e_2}$:
        let  $ t = \mthpair{t_1}{t_2} $  where $ t_i $ matches type of $ e_i$;
          insertTuple($t_1, e_1$) ; insertTuple($t_2, e_2$) \\
 \tb $e_1 - e_2$: insertTuple($t, e_1$) ;  deleteTuple($t, e_2$) \\
\tb   $e_1 . e_2$:
       let $T$ be the common sig-type that joins $e_1$ and $e_2$; \\
  \tb  \tb   \tb  if  $ T$ is the type of $e_1$ then
  for some $ a $ in $\eval(e_1, I, I')$, insertTuple($\mthpair{a}{t}, e_2$) \\
  \tb \tb \tb elseif $ T $ is the type of $ e_2 $ then
  for some $ a $ in $\eval(e_2, I, I')$, insertTuple($\mthpair{t}{a}, e_1$) \\
  \tb \tb \tb else \textbf{choose} $ a : T$ ; set
  $t_1 =  \mthpair{s_1}{a}$ and set  $t_2 =\mthpair{a}{s_2}$; \\
  \tb \tb \tb \tb insertTuple($t_1, e_1$)  ; insertTuple($t_2, e_2$) \\
\tb $\mthtc{(e_1)}$: insertTuple($t, e_1$)
\end{quote}

\begin{quote}
\textbf{procedure } deleteTuple($t:$ tuple, \; $e$: expression) \{ \\
\tb match $e$:  \\
  \tb atom $a$: if $a = t $ then FAIL else RETURN \\
  \tb immutable relation $r$: if $t \in r$ then FAIL else RETURN \\
  \tb mutable relation $r$: if $t$ has been previously inserted into $r$
  then FAIL \\
  \tb \tb \tb else delete $t$ from the table $r$ in $J'$ \\
  \tb   $e_1 \mthunion e_2$:  deleteTuple($t, e_1$) ; deleteTuple($t, e_2$) \\
  \tb $e_1 \mthint e_2$:  \textbf{choose} some $ e_i $ ;  deleteTuple($t,
  e_i$) \\
  \tb $\sim e$: deleteTuple($~t, e)$ \\
  \tb $\mthpair{e_1}{e_2}$:
  let  $ t = \mthpair{t_1}{t_2} $  where $ t_i $ matches type of $ e_i$;
  \textbf{choose} some $e_i$; deleteTuple($t_i, e_i$)  \\
  \tb $e_1 - e_2$:  \textbf{choose:} deleteTuple($t, e_1$) or insertTuple($t, e_2$) \\
  \tb   $e_1 . e_2$:
  let $T$ be the common sig-type that joins $e_1$ and $e_2$; \\
  \tb  \tb   \tb  if  $ T$ is the type of $e_1$ then
  for each $ a $ in $\eval(e_1, I, I')$, deleteTuple($\mthpair{a}{t}, e_2$) \\
  \tb \tb \tb elseif $ T $ is the type of $ e_2 $ then
  for each $ a $ in $\eval(e_2, I, I')$, deleteTuple($\mthpair{t}{a}, e_1$) \\
  \tb \tb \tb else for each $ a : T$ such that for some $s_1, s_2$, \\
  \tb \tb \tb \tb \tb   $ \mthpair{s_1}{a} = t_1$ is in $e_1$ and 
  $\mthpair{a}{s_2} = t_2$ is in $e_2$ and $ t_1 . t_2 = t$; \\
  \tb \tb \tb \tb \tb \textbf{choose} $e_i$ then deleteTuple($t_i, e_i$) \\
  \tb $\mthtc{(e_1)}$: for each $(x, y_1), (y_1, y_2), \ldots, (y_n, y)$ such that
  $t = (x, y)$  and each pair is in $ e_1$ \\
  \tb \tb \tb \textbf{choose} some pair  $(y_i, y_{i+1})$; 
  deleteTuple($\mthpair{y_i}{y_{i+1}}, e_1$)
\end{quote}

\subsection{Proof of correctness} \label{correctness}

\paragraph{Proof of \refthm{thm-main}}
\label{sec:proof-refthmthm-main}
\refthm{thm-main} follows from the following lemma about \thealg.

\begin{lemma}
Let $p$ be a predicate; let \thealg be
  the non-deterministic procedure constructed from $p$ by \refdef{def:thealg}.
  Then for every instance $\inst$ and binding $\env$ for the parameters
  of $p$:
  \begin{enumerate}
  \item %
        Every computation of \thealg  terminates on  $\inst$ under
        $\env$, and if \thealg returns an instance
        $\inst'$, we have
        $(\inst, \inst') \in \semenv{p}$;
      \item %
        If there is an instance $\inst'$ such that %
        $(\inst, \inst') \in \sem{p}(\env)$ then \thealg will not fail.
\end{enumerate}
\end{lemma}
\begin{proof}[Proof of the lemma]
  For the first claim, first note that algorithm \helperalg proceeds by
  primitive recursion over the body of the predicates and algorithms
  insertTuple and deleteTuple proceed by primitive recursion over the
  body of expressions.  So it suffices to argue that the iteration until
  fixed point in algorithm \thealg always terminates.  But this follows
  from the fact that we never add or delete the same tuple from a given
  relation and the total size of the domain we work with never changes.
  It is easy to see that when \thealg halts without failure it is the
  case that the body of the predicate has been satisfied.

To establish the second claim we start with a definition.   Given instances
$\inst$ and $\inst'$ let us say that instance $J$ is an %
\emph{$(I, I')$-approximation} if
$I - J \subseteq I - I' $ and
$ J - I  \subseteq I' - I.$
We abuse notation slightly here: these calculations are done on a
per-relation basis. %
Intuitively $J$ is an $(I, I')$-approximation if $J$ can be obtained
from $I$ by making \emph{some} of the inserts and deletes that transform
$I$ into $I'$.  Note that $I$ is an $(I, I')$-approximation, as is $I'$.
Now the second claim follows from the fact that, for initial instance
$I$ and chosen $I'$ with $(\inst, \inst') \in \sem{p}(\env)$,
whenever algorithm \helperalg is called (by \thealg) when the current
value of the poststate is an $(I, I')$-approximation then there is a
computation of \helperalg that (i) does not fail, and (ii) updates the
poststate so that it still is an $(I, I')$-approximation.  In
particular \thealg will never fail.
\end{proof}

\paragraph{Complexity}
There is nothing interesting that can be said about the run-time
complexity of $\cmd{p}$ since it depends on the nature of the predicate
$p$, and $p$ can be an arbitrary predicate.  On the other hand it is natural to ask about
the complexity of $\cmd{}$ itself.  In other words, what is the
running time \emph{of Alchemy's code generation algorithm?}  Since
$\cmd{p}$ comprises a backtracking wrapper around the algorithm \thealg
the question is essentially the same as asking:   what is the complexity
of building the text of algorithm \thealg from the text of predicate
$p$?    It is easy to see that this is linear in $p$.  Note in
particular that the procedures insertTuple and deleteTuple do not depend
on $p$ at all.

\section{Related Work}
\label{sec:relwork}

For an extensive discussion of previous research relevant to the Alchemy
project itself we refer the reader to the related work section in
\cite{kdfy:alchemy-trans-alloy-spec-impl}.  The relationship of the
present paper to the previous  work on Alchemy is as follows.  In
\cite{kdfy:alchemy-trans-alloy-spec-impl} we did not handle the
relational difference operator, we did not treat Skolemization, and our
correctness result was only for a subset of Alloy predicates (those
admitting ``homogeneous'' implementations as defined there).  But most
importantly, the treatment of when relation names were evaluated in the
pre-state and when in the post-state was ad-hoc: in the current paper
this important semantic decision rests on the secure foundations of the
work in \cite{gdfk:op-sem-alloy}.  This allows us to prove a true
soundness and completeness theorem (\refthm{thm-main}) for our
code-generation algorithm.

\vspace*{-5pt}
\bibliographystyle{eptcs}
\bibliography{rule}

\end{document}